\documentclass[aip,preprint]{revtex4-2}
\usepackage[cp1251]{inputenc}
\usepackage[T2A]{fontenc}
\usepackage[english]{babel}
\usepackage{amssymb,latexsym,amsmath,amscd}
\usepackage{graphicx,color,framed}
\usepackage{bm}
\usepackage{tikz}
\usepackage{xcolor}
\usepackage{siunitx}
\usepackage{empheq}
\usepackage{xr}
\usepackage{microtype}
\usepackage{amsfonts}
\usepackage[normalem]{ulem}
\graphicspath{{figures/}}

\makeatletter
\newcommand*{\addFileDependency}[1]{% argument=file name and extension
  \typeout{(#1)}
  \@addtofilelist{#1}
  \IfFileExists{#1}{}{\typeout{No file #1.}}
}
\makeatother

\begin{document}
\title{Statistical field theory of mechanical stresses in Coulomb fluids. Noether's theorem {\sl vs} General covariant approach}
\author{\firstname{Petr E.} \surname{Brandyshev}}
%\homepage[]{Your web page}
%\thanks{}
%\altaffiliation{}
\affiliation{Laboratory of Computational Physics, HSE University, Tallinskaya st. 34, 123458 Moscow, Russia}
\author{\firstname{Yury A.} \surname{Budkov}}
\email[]{ybudkov@hse.ru}
%\homepage[]{Your web page}
%\thanks{}
%\altaffiliation{}
\affiliation{Laboratory of Computational Physics, HSE University, Tallinskaya st. 34, 123458 Moscow, Russia}
\affiliation{Laboratory of Multiscale Modeling of Molecular Systems, G.A. Krestov Institute of Solution Chemistry of the Russian Academy of Sciences, 153045, Akademicheskaya st. 1, Ivanovo, Russia}

\begin{abstract}
In this paper, we introduce a statistical field theory that describes the macroscopic mechanical forces in inhomogeneous Coulomb fluids. Our approach employs the generalization of Noether's first theorem for the case of fluctuating order parameter, to calculate the stress tensor for Coulomb fluids. This tensor encompasses the mean-field stress tensor and the fluctuation corrections derived through the one-loop approximation. The correction for fluctuations includes a term that accounts for the thermal fluctuations of the local electrostatic potential and field in the vicinity of the mean-field configuration. This {\sl correlation stress tensor} determines how electrostatic correlation affects local stresses in a nonuniform Coulomb fluid. We also use previously formulated general covariant methodology [P.E. Brandyshev and Yu.A. Budkov, J. Chem. Phys. 158, 174114 (2023)] in conjunction with a functional Legendre transformation method and derive within it the same total stress tensor. We would like to emphasize that our general approaches are applicable not only to Coulomb fluids but also to nonionic simple or complex fluids, for which the field-theoretic Hamiltonian is known as a functional of the relevant scalar order parameters.
\end{abstract}

\maketitle

\section{Introduction}
Coulomb fluids, such as plasma, electrolyte solutions, molten salts, and room temperature ionic liquids, have become a popular topic among chemical engineers and researchers due to their use in various applications~\cite{naji2013perspective}. These applications include lipid and ion-exchange membranes, biomacromolecules, colloids, batteries, fuel cells, and supercapacitors, all of which involve Coulomb fluids interacting with charged surfaces or being confined in charged nanopores. However, the inhomogeneity of the ionic fluid violates its local electrical neutrality, which requires the use of numerical solutions of the self-consistent field equation (Poisson-Boltzmann equation or its modifications) for the electrostatic potential with appropriate boundary conditions~\cite{blossey2023poisson,budkov2022modified,budkov2021electric,borukhov1997steric,abrashkin2007dipolar,avni2020charge,podgornik2018general_,kornyshev2007double,iglivc2010excluded,bazant2011double,slavchov2014quadrupole}.

In practical applications of Coulomb fluids confined in nanosized pores of varying geometries, it is necessary to calculate the mechanical stresses described by the stress tensor in addition to the concentration and electrostatic potential profiles. This knowledge of the local stress tensor, which is consistent with a certain self-consistent field equation, enables the calculation of physical quantities such as solvation pressure and shear stresses, which are useful for estimating the deformation of pore materials (in batteries and supercapacitors, for instance)~\cite{kolesnikov2022electrosorption,kolesnikov2021models}. Furthermore, the stress tensor can be used to calculate the macroscopic force acting on the charged macroscopic conductor or dielectric immersed in the Coulomb fluid~\cite{kolesnikov2022electrosorption,neu1999wall,trizac1999long,de2020continuum,ruixuan2023electrostatic}. Therefore, a first-principle approach to derive the stress tensor of inhomogeneous Coulomb fluids would be beneficial for practical application purposes.

Recent progress has been made in this field, as demonstrated by two recent articles~\cite{budkov2022modified,budkov2023macroscopic}. In the first~\cite{budkov2022modified}, Budkov and Kolesnikov applied Noether's (first) theorem~\cite{hermann2022noether,noether1971invariant} to the grand thermodynamic potential of a Coulomb fluid as a functional of the electrostatic potential. They established a conservation law for the symmetric stress tensor, $\sigma_{ik}$, which represents the local mechanical equilibrium condition. This tensor is composed of two terms: the electrostatic Maxwell stress tensor, which is related to the local electric field, and the hydrostatic isotropic stress tensor, which is determined by the local osmotic pressure of the ions. The authors extended this equilibrium condition to include cases where external potential forces act on the ions. They also derived a general analytical expression for the electrostatic disjoining pressure of a Coulomb fluid confined in a charged nanopore slit, which goes beyond the conventional DLVO expression to include different reference models of fluid. In another study~\cite{budkov2023macroscopic}, Budkov and Kalikin presented a self-consistent field theory of macroscopic forces in inhomogeneous flexible chain polyelectrolyte solutions. The authors derived an analytical expression for a stress tensor by subjecting the system to a small dilation and considering the extremum of the grand thermodynamic potential. This stress tensor includes the previously mentioned hydrostatic and Maxwell stress tensors, as well as a conformational stress tensor generated by the conformational Lifshitz entropy of flexible polymer chains. The authors applied their theory to the investigation of polyelectrolyte solution constrained in a conducting slit nanopore and observed anomalies in the disjoining pressure and electric differential capacitance at small pore thicknesses.

It is important to mention several earlier works where the authors explored the derivation and explicit evaluation of electrostatic normal stress in Coulomb fluids confined in slit-like pores at both weak-coupling and strong-coupling limits~\cite{moreira2001binding,dean2003field,jho2008strong,buyukdagli2023impact}. Moreira and Netz utilized Monte-Carlo simulations and statistical field theory, specifically on the level of one-loop approximation and strong coupling theory, to investigate the behavior of highly charged plates in the presence of multivalent counterions. Their study revealed a novel unbinding transition at the equilibrium plate separation where the pressure changes from attractive to repulsive. In another study~\cite{dean2003field}, Dean and Horgan proposed a contact value theorem for Coulomb fluids in planar or film-like geometries using a Hamiltonian field theoretic representation of the system. Furthermore, Jho et al. examined the strong-coupling electrostatic interaction between two like-charged nanoparticles through the use of strong coupling theory and Monte-Carlo simulations. In recent paper by Buyukdagli~\cite{buyukdagli2023impact}, a contact-value identity was derived by considering the field-theoretic partition function of an electrolyte confined between two anionic membranes. This identity holds true for a wide range of intramolecular solute structures and electrostatic coupling strengths.

Despite the progress made in deriving the local stress tensor within the mean-field theory, and normal electrostatic stress in slit-like pores even beyond mean-field theory, it is still unclear how to compute all components of stress tensor beyond the mean-field approximation.

The fluctuation corrections to the mean-field stress tensor should occur when we consider the fluctuation corrections for the thermodynamic potential. The fluctuation corrections for the mean-field approximation of the thermodynamic potential are always non-local functionals~\cite{netz2000beyond,lau2008fluctuation}, even on a one-loop correction level, which makes the use of Noether's theorem, formulated for local functionals~\cite{noether1971invariant}, quite problematic. Thus, it would be valuable to have a generalization of the Noether's theorem formalism~\cite{noether1971invariant} for the case of fluctuating order parameters when dealing with the formulation in terms of a functional integral over the fluctuating order parameters. On the other hand, an alternative approach to Noether's theorem for this case could be the recently proposed by us in the paper~\cite{brandyshev2023noether} general covariant approach. Our approach is based on Noether's second theorem, which allows us to derive the symmetric stress tensor for any model of an inhomogeneous liquid as a functional derivative of a grand thermodynamic potential with respect to the metric tensor. It is important to note that the general covariant approach does not offer any advantages over Noether's first theorem when it comes to local functionals. However, it can be more advantageous for nonlocal functionals~\cite{budkov2023variational}. We have applied this approach to several phenomenological nonlocal models of inhomogeneous Coulomb fluids, such as the Cahn-Hilliard-like model~\cite{blossey2017structural,vasileva2023theory}, the Bazant-Storey-Kornyshev model~\cite{bazant2011double}, and the Maggs-Podgornik-Blossey model~\cite{blossey2017structural}, and have obtained the corresponding phenomenological stress tensors. It should be noted that this method is similar to the one used by D. Hilbert in the general theory of relativity to derive the energy-momentum tensor from the action functional~\cite{landau1971classical,earman1978einstein}.

The present paper proposes two methods for deriving the stress tensor of inhomogeneous Coulomb fluids.
The first method is based on the application of Noether's first theorem to the grand partition function, which is presented in functional integral form. The second one is based on the aforementioned general covariant approach in combination with the functional Legendre transform, to obtain fluctuation corrections to the mean-field approximation of the stress tensor for a Coulomb fluid.

\section{Functional Legendre transform}
We start from the generating functional as the following functional integral~\cite{budkov2022modified} over the fluctuating electrostatic potential $\phi$ with the auxiliary function $\rho(\bold{r})$:
\begin{equation}\label{1.1}
\Xi[\rho]=\int \frac{\mathcal{D}\phi}{C_0} \exp\bigg(-\frac{\mathcal{H}[\phi]+(\rho\phi)}{k_{B}T}\bigg)  
\end{equation}
where
\begin{equation}
\label{H}
\mathcal{H}[\phi]=\int d\bold{r} \bigg[\frac{\varepsilon}{2}(\nabla\phi)^2-P(\{\bar{\mu}_{\alpha}\})\bigg],
\end{equation}
is the field-theoretic Hamiltonian obtained in \cite{budkov2022modified} and
\begin{equation}
C_0=\exp\bigg[-\frac{1}{2}tr \ln\bigg(-\frac{\varepsilon}{k_{B}T}\Delta\bigg)\bigg],
\end{equation}
is the normalizing multiplayer of the Gaussian measure; $P(\{\bar{\mu}_{\alpha}\})$ is the pressure of the reference fluid system (see ref.~\cite{budkov2022modified}) dependent on the "shifted" chemical potentials $\bar{\mu}_{\alpha}=\mu_{\alpha}+iq_{\alpha}\phi-u_{\alpha}$; $k_B$ is the Boltzmann constant, $T$ is the temperature, $u_{\alpha}$ are the external potentials, $q_{\alpha}$ is the electric charge of the ion of $\alpha$th kind; $\Delta=\nabla^2$ is the Laplace operator. We have also introduced the short-hand notation
\begin{equation}
(\rho\phi)=\int d\bold{r} \rho(\bold{r})\phi(\bold{r}).
\end{equation}
At $\rho(\bold{r})=0$ the generating functional (\ref{1.1}) transforms into the grand partition function of the Coulomb fluid obtained in ref.~\cite{budkov2022modified}.  We emphasize that the present study only deals with the simplest model of the Coulomb fluid (Model I, as classified in~\cite{budkov2022modified}). In other words, this model does not account for the static polarizabilities and permanent dipole moments of the ions and does not explicitly consider the polar solvent. Nevertheless, a generalization of the theory for the latter cases (Models II and III) can be performed directly. We also emphasize that our study focuses on inhomogeneous Coulomb fluids, which involve the interaction between ions and surface external charges. These charges, although not included in the field-theoretic Hamiltonian, can be taken into account by incorporating them into the boundary conditions for the self-consistent field equations. Furthermore, the spatial inhomogeneity of the Coulomb fluid can also be attributed to the presence of external potential fields characterized by potentials $u_{\alpha}(\bold{r})$.

The generating functional can be rewritten as follows
\begin{equation}
\Xi[\rho]=\exp\bigg(-\frac{W[\rho]}{k_{B}T}\bigg),
\end{equation}
where $W[\rho]$ is some functional of the auxiliary function $\rho(\bold{r})$. Thus, the expectation value can be obtained from the equation
\begin{equation}
\label{1.2}
 i\varphi(\bold{r})=\frac{\delta W[\rho]}{\delta \rho(\bold{r})}.
\end{equation}
Note that at $\rho=0$ we have $\varphi(\bold{r})=-i\left<\phi(\bold{r})\right>$, where $\left<(..)\right>$ means average over the Gibbs statistics of the Coulomb fluid with the field-theoretical Hamiltonian (\ref{H}). The effective grand thermodynamic potential (GTP) can be derived from the following functional Legendre transform~\cite{weinberg1995quantum_2,netz2000beyond,lau2008fluctuation}
\begin{equation}\label{1.8}
\Omega[\phi]=W[\rho]-(\rho\phi).
\end{equation}
Therefore,
\begin{equation}
\frac{\delta\Omega[\phi]}{\delta\phi(\bold{r})}=\int d\bold{r}'
\frac{\delta W[\rho]}{\delta\rho(\bold{r}')}\frac{\delta\rho(\bold{r}')}{\delta\phi(\bold{r})}-\int d\bold{r}'\frac{\delta\rho(\bold{r}')}{\delta\phi(\bold{r})}\phi(\bold{r}')-\rho(\bold{r})
\end{equation}
and with account of eq. (\ref{1.2}), we get 
\begin{equation}
\label{1.9}
\frac{\delta\Omega[i\varphi]}{\delta\varphi(\bold{r})}=-i\rho(\bold{r}).
\end{equation}
The mean-field potential (or saddle-point), $\psi(\bold{r})$, can be obtained from the following Euler-Lagrange (EL) equation
\begin{equation}\label{1.3}
\frac{\delta\mathcal{H}[i\psi]}{\delta\psi(\bold{r})}=-i\rho(\bold{r}).
\end{equation}
Introducing the fluctuation, $\eta(\bold{r})$, near the mean-field potential by
\begin{equation}
\phi=i\psi+\eta,
\end{equation}
we can expand $\mathcal{H}$ in a functional series in $\eta$ 
\begin{equation}
\frac{\mathcal{H}[i\psi+\eta]}{k_{B}T}=\frac{\mathcal{H}[i\psi]}{k_{B}T}+
\frac{\varepsilon}{2k_{B}T}\int d\bold{r} \eta(\bold{r})\bigg(-\Delta+\varkappa^2(\bold{r})\bigg)\eta(\bold{r}),
\end{equation}
where
\begin{equation}
\label{varkappa}
\varkappa^2(\bold{r})=U(\psi(\bold{r}))=\frac{1}{\varepsilon}\sum\limits_{\alpha\gamma}{q_\alpha q_\gamma}\frac{\partial^{2}P}{\partial\bar{\mu}_\alpha \partial\bar{\mu}_\gamma},
\end{equation}
and calculating the Gaussian integral (\ref{1.1}) over the fluctuations $\eta$, we arrive at
\begin{equation}\label{1.4}
W[\rho]=\mathcal{H}[i\psi]+\frac{\Omega_1[\psi]}{2\beta}+i(\rho\psi)+O\bigg(\frac{1}{\beta^2}\bigg).
\end{equation}
where the functional dependence of $\psi$ on $\rho$ is determined by eq. (\ref{1.3}), $\beta=1/k_{B}T$ and the following notation
\begin{equation}
\Omega_1[\psi]=tr\ln\bigg(\frac{-\Delta+U}{-\Delta}\bigg)
\end{equation}
is introduced. Symbol $tr(..)$ denotes the trace of operator. We would like to note that $1/\beta$ plays the role of the Planck's constant $\hbar$ in quantum field theory~\cite{weinberg1995quantum_2}. The occurrence of the multiplayer $1/\beta=k_{B}T$ in the second term on the right-hand side of equation (\ref{1.4}) reflects the fact that this term describes the contribution of thermal fluctuations of the electrostatic potential near its mean value, $\varphi$. Note that in the present work we hold only the first-order terms on $1/\beta$.

Using eqs. (\ref{1.2}) and (\ref{1.4}), we can get
\begin{equation}\label{1.10}
i\varphi(\bold{r})=\int d\bold{r}'\bigg(
\frac{\delta\mathcal{H}[i\psi]}{\delta \psi(\bold{r}')}+i\rho(\bold{r}')\bigg)\frac{\delta\psi(\bold{r}')}{\delta \rho(\bold{r})}+i\psi(\bold{r})+\frac{i\chi(\bold{r})}{2\beta},
\end{equation}
where
\begin{equation}\label{1.13}
i\chi(\bold{r})=\int d\bold{r}'
\frac{\delta\Omega_1[\psi]}{\delta\psi(\bold{r}')}\frac{\delta\psi(\bold{r}')}{\delta \rho(\bold{r})}.
\end{equation}
Taking into account eq. (\ref{1.3}) as it follows from eq. (\ref{1.10}), we obtain
\begin{equation}
 \psi(\bold{r})=
\varphi(\bold{r})-\frac{\chi(\bold{r})}{2\beta}.  
\end{equation}
Substituting this into eq. (\ref{1.4}) and expanding it in a series in $\chi$, we get
\begin{equation}\label{}
\begin{aligned}
W[\rho]=\mathcal{H}[i\varphi]+
i(\rho\varphi)-\int d\bold{r}\bigg(
\frac{\delta\mathcal{H}[i\psi]}{\delta \psi(\bold{r})}+i\rho(\bold{r})\bigg)\frac{\chi(\bold{r})}{2\beta}
+\frac{\Omega_1[\varphi]}{2\beta}.    
\end{aligned}
\end{equation}
and using eq. (\ref{1.3}) again, we arrive at
\begin{equation}\label{}
\begin{aligned}
W[\rho]=\mathcal{H}[i\varphi]+
\frac{\Omega_1[\varphi]}{2\beta}+i(\rho\varphi).    
\end{aligned}
\end{equation}
Then performing the Legendre transform (\ref{1.8}),  we eventually obtain
\begin{equation}
\label{GTP}
\Omega[i\varphi]=\mathcal{H}[i\varphi]
+\frac{\Omega_1[\varphi]}{2\beta}.
\end{equation}
Eq. (\ref{1.3}) at $\rho=0$ can be written as
\begin{equation}\label{1.12}
\varepsilon\Delta\psi=-\sum_{\alpha}q_{\alpha}\bar{c}_{\alpha},
\end{equation}
where we took into account that ${\partial P(\psi)}/{\partial\psi}=-\sum_{\alpha}q_{\alpha}\bar{c}_{\alpha}$ and $\partial{P}/\partial{\bar{\mu}_{\alpha}}=\bar{c}_{\alpha}$. Eq. (\ref{1.12}) is nothing more than the modified Poisson-Boltzmann equation~\cite{budkov2022modified}. 

Variation of $\Omega_1$ is derived from the expression
\begin{equation}\label{1.15}
\frac{\delta \Omega_1[\varphi]}{\delta\varphi(\bold{r})}=\beta\varepsilon\frac{\partial U(\varphi)}{\partial\varphi(\bold{r})}G(\bold{r},\bold{r}|\varphi),
\end{equation}
where the Green's function $G(\bold{r},\bold{r}'|\varphi)$ is determined by equation
\begin{equation}\label{1.16}
\varepsilon\beta\bigg(-\Delta+U(\varphi)\bigg)G(\bold{r},\bold{r}'|\varphi)=\delta(\bold{r}-\bold{r}').
\end{equation}
Thus the EL equations (\ref{1.9}) at $\rho=0$ has the form
\begin{equation}\label{1.11}
-\frac{\partial P}{\partial\varphi}+\varepsilon\Delta\varphi+\frac{\varepsilon}{2}\frac{\partial U(\varphi)}{\partial\varphi(\bold{r})}G(\bold{r},\bold{r}|\varphi)=0.
\end{equation}
Then using the expansion (\ref{1.10}) and holding first-order terms on $1/\beta$, as mentioned above, we can write eq. (\ref{1.11}) with account of eq. (\ref{1.12}) as follows
\begin{equation}\label{1.14}
\bigg(-\Delta+U(\psi)\bigg)\frac{\chi(\bold{r})}{\beta}-\frac{\partial U(\psi)}{\partial\psi(\bold{r})}G(\bold{r},\bold{r}|\psi)=0.
\end{equation}
Now, using eqs. (\ref{1.3}) and (\ref{1.16}), we can show that
\begin{equation}\label{1.17}
\frac{\delta\psi(\bold{r}')}{\delta\rho(\bold{r})}=i\beta G(\bold{r},\bold{r}'|\psi).
\end{equation}
Thus, substituting eqs. (\ref{1.15}) and (\ref{1.17}) into eq. (\ref{1.13}), we get
\begin{equation}\label{1.18}
\chi(\bold{r})=\varepsilon\beta^2\int d\bold{r}'G(\bold{r},\bold{r}'|\psi)
G(\bold{r}',\bold{r}'|\psi)\frac{\partial U(\psi)}{\partial\psi(\bold{r}')}.
\end{equation}
By examining eq. (\ref{1.16}), we can conclude that $G(\bold{r},\bold{r}'|\psi)$ is a first-order value on $1/\beta$. As a result, it is expected that $\chi$ would also be a zero-order value on $1/\beta$, corroborating the expansion given in eq. (\ref{1.10}). Furthermore, we can verify that eq. (\ref{1.18}) is indeed a solution to the EL equation (\ref{1.14}).

\section{Stress tensor. Noether's theorem}
In this section, we would like to formulate an approach based on a generalization of Noether's first theorem \cite{noether1971invariant,hermann2022noether,budkov2022modified} to derive the stress tensor of Coulomb fluids from the grand partition function presented in the functional integral form taking into account electrostatic field thermal fluctuations. The grand partition function is
\begin{equation}\label{}
\Xi=\int \frac{\mathcal{D}\phi}{C_0} \exp\bigg(-\frac{\mathcal{H}[\phi])}{k_{B}T}\bigg),
\end{equation}
where the functional $\mathcal{H}$ is determined by eq. (\ref{H}). First, let us consider the case of $u_{\alpha}(\bold{r})=0$.

Let us perform the global infinitesimal shift transformation of the coordinates
\begin{equation}\label{}
x'_{k}=x_{k}+h_{k},
\end{equation}
under which the grand partition function have to be invariant, i.e.
\begin{equation}\label{}
\delta\ln\Xi=-\frac{1}{k_{B}T}\bigg\langle\delta\mathcal{H}[\phi]\bigg\rangle=0.
\end{equation}
Then, after some algebra~\cite{hermann2022noether,brandyshev2023noether}, we obtain
\begin{equation}\label{}
\partial_{i}\bigg\langle\hat{\sigma}_{ik}h_{k}\bigg\rangle=\bigg\langle \frac{\delta\mathcal{H}[\phi]}{\delta\phi}h_{k}\partial_{k}\phi \bigg\rangle,
\end{equation}
where we have introduced the stress tensor determined on the fluctuating random order parameter (electrostatic potential)
\begin{equation}\label{5_1}
\hat{\sigma}_{ik}=
\frac{\varepsilon}{2}\delta_{ik}\partial_j\phi\partial_j\phi
-\varepsilon\partial_i\phi\partial_k\phi-P\delta_{ik}.
\end{equation}
Since $h_{k}$ is the arbitrary constant infinitesimal vector, we can get
\begin{equation}\label{}
\partial_{i}\langle\hat{\sigma}_{ik}\rangle=\bigg\langle\frac{\delta\mathcal{H}[\phi]}{\delta\phi}\partial_{k}\phi\bigg\rangle,
\end{equation}
where the expectation value is
\begin{equation}\label{1_1}
\sigma_{ik}=\bigg\langle\hat{\sigma}_{ik}\bigg\rangle=\frac{1}{\Xi}\int \frac{\mathcal{D}\phi}{ C_0}\hat{\sigma}_{ik}\exp\bigg(-\frac{\mathcal{H}[\phi])}{k_{B}T}\bigg).
\end{equation}

Further, taking into account that~\cite{buyukdagli2020schwinger}
\begin{equation}
\bigg\langle\frac{\delta\mathcal{H}[\phi]}{\delta\phi(\bold{r})}\partial_{k}\phi(\bold{r})\bigg\rangle= \partial_{k}\frac{\delta\phi(\bold{r})}{\delta\phi(\bold{r})}=\partial_{k}\delta(\bold0)= \lim_{\bold{r}'\rightarrow \bold{r}}\bigg(\partial_{k}+\partial_{k}'\bigg)\delta(\bold{r}-\bold{r}')=0,
\end{equation}
we arrive at Noether's first theorem, generalized for the case of fluctuating order parameter
\begin{equation}\label{Noether}
\partial_{i}\langle\hat{\sigma}_{ik}\rangle=0.
\end{equation}
The latter expression, which is the local mechanical equilibrium condition with the average stress tensor $\langle\hat{\sigma}_{ik}\rangle$ represents another important result of this paper.

In the case that external fields occur ($u_{\alpha}(\bold{r})\neq 0$), the same calculations lead to the following mechanical equilibrium condition
\begin{equation}\label{Noether_2}
\partial_{i}\langle\hat{\sigma}_{ik}\rangle-\sum\limits_{\alpha}\langle\hat{c}_{\alpha}\rangle\partial_{k}u_{\alpha}=0,
\end{equation}
where $\hat{c}_{\alpha}(\bold{r})=\partial P/\partial\bar{\mu}_{\alpha}$ is the microscopic fluctuating ionic concentrations.

Now, let us estimate the average stress tensor, $\sigma_{ik}=\left<\hat{\sigma}_{ik}\right>$. For this purpose, we introduce the fluctuation, $\eta(\bold{r})$, near the mean-field potential by
\begin{equation}\label{6}
\phi=i\psi+\eta.
\end{equation}
Taking into account that fluctuation $\eta$ is the value of order $(k_{B}T)^{\frac{1}{2}}$, substituting (\ref{6}) into (\ref{5_1}), we can show that the fluctuating stress tensor has the form with accuracy up to terms of order $k_BT$
\begin{equation}\label{2*}
\begin{aligned}
\hat{\sigma}_{ik}=
\sigma^{(MF)}_{ik}-\delta_{ik}\sum\limits_{\lambda}\mu_{\lambda}^{(1)}\bar{c}_{\lambda}+i\varepsilon\bigg(\delta_{ik}\partial_j\eta\partial_j\psi-\partial_i\eta\partial_k\psi-
\partial_i\psi\partial_k\eta\bigg)+i\eta\delta_{ik} \frac{\partial P}{\partial \psi}\\
+\frac{\varepsilon}{2}U(\psi)\eta^2\delta_{ik}+
\frac{\varepsilon}{2}\delta_{ik}\partial_j\eta\partial_j\eta-\varepsilon\partial_i\eta\partial_k\eta,
\end{aligned}
\end{equation}
where we took into account that $P(\{\mu_{\alpha}-q_{\alpha}\psi \})=P(\{\mu_{\alpha}^{(0)}-q_{\alpha}\psi \})+\sum_{\lambda}\mu_{\lambda}^{(1)}\bar{c}_{\lambda}$. Note that the fluctuation corrections $\mu_{\alpha}^{(1)}$ for the chemical potentials of the mean-field should be calculated for each individual case~\cite{netz2000beyond,moreira2001binding,lau2008fluctuation}.

As it was obtained in ref.~\cite{budkov2022modified}, the mean-field approximation for the stress tensor is
\begin{equation}
\label{sigma_MF}
\sigma^{(MF)}_{ik}=\varepsilon\partial_i\psi\partial_k\psi-
\frac{\varepsilon}{2}\delta_{ik}\partial_j\psi\partial_j\psi-P\delta_{ik}.
\end{equation}
In what follows, utilizing eq. (\ref{Noether}), we will calculate the fluctuation corrections to the mean-field approximation (\ref{sigma_MF}). 

Thus, to take into account corrections to the stress tensor up to the first order on $k_{B}T$, we have to restrict ourselves to an expansion of $\mathcal{H}$ up to the third order on $\eta$. Then, using EL equation
\begin{equation}\label{4}
\frac{\delta\mathcal{H}[i\psi]}{\delta\psi}=0,
\end{equation}
we can expand $\mathcal{H}$ in a functional series in $\eta$ as follows
\begin{equation}
\begin{aligned}
\frac{\mathcal{H}[i\psi+\eta]}{k_{B}T}=\frac{\mathcal{H}[i\psi]}{k_{B}T}+
\frac{\varepsilon}{2k_{B}T}\int d\bold{r} \eta(\bold{r})\bigg(-\Delta+U(\psi)\bigg)\eta(\bold{r})\\
-\frac{i\varepsilon}{6k_BT}\int d\bold{r}\frac{\partial U(\psi)}{\partial\psi}\eta^3(\bold{r})+\frac{O(\eta^4)}{k_BT}.
\end{aligned}
\end{equation}
Then, we have
\begin{equation}\label{3*}
\begin{aligned}
e^{-\beta\mathcal{H}[i\psi+\eta]}=e^{-\beta\mathcal{H}[i\psi]}e^{-\frac{1}{2}(\eta G^{-1}\eta)}
\bigg(1+\frac{i\varepsilon\beta}{6}\int d\bold{r}\frac{\partial U(\psi)}{\partial\psi}\eta^3(\bold{r})+O\bigg(\frac{1}{\beta}\bigg)\bigg),
\end{aligned}
\end{equation}
where we have introduced the following short-hand notation
\begin{equation}\label{}
(\eta G^{-1}\eta)=\int d\bold{r}\int d\bold{r}' \eta(\bold{r}) G^{-1}(\bold{r},\bold{r}'|\psi)\eta(\bold{r}').
\end{equation}

Substituting (\ref{2*}) and (\ref{3*}) into (\ref{1_1}) and integrating on $\eta$ we get the stress tensor with accuracy up to terms of order $1/\beta$
\begin{equation}
\label{12}
\begin{aligned}
 \sigma_{ik}=\sigma^{(0)}_{ik}+\frac{\varepsilon}{2\beta}\bigg(\partial_i\chi\partial_k\psi+\partial_i\psi\partial_k\chi-
\delta_{ik}\partial_j\chi\partial_j\psi\bigg)-\delta_{ik} \frac{\chi}{2\beta}\frac{\partial P}{\partial \psi}+\sigma^{(1)}_{ij},
\end{aligned}
\end{equation}
where
\begin{equation}\label{7_2}
\sigma^{(0)}_{ik}=\sigma^{(MF)}_{ik}-\delta_{ik}\sum\limits_{\lambda}\mu_{\lambda}^{(1)}\bar{c}_{\lambda},
\end{equation}
\begin{equation}\label{8_2}
\begin{aligned}
\sigma^{(1)}(\bold{r})=\sigma^{(cor)}_{ij}(\bold{r})=\frac{\varepsilon}{2}\left(U(\psi)G(\bold{r},\bold{r}|\psi)
+\mathcal{D}_{kk}(\bold{r})\right)\delta_{ij}-\varepsilon \mathcal{D}_{ij}(\bold{r}),
\end{aligned}
\end{equation}
and following short-hand notations
\begin{equation}
\mathcal{D}_{ij}(\bold{r})= \lim_{\bold{r}'\rightarrow \bold{r}}
\partial_i\partial_j' G(\bold{r},\bold{r}'|\psi),
\end{equation}
\begin{equation}\label{16}
\chi(\bold{r})=\frac{\varepsilon\beta^2}{3}\int d\bold{r}'\frac{\partial{U(\psi(\bold{r}'))}}{\partial\psi(\bold{r}')}\bigg\langle\eta(\bold{r})\eta^3(\bold{r}') \bigg\rangle_{0}
\end{equation}
have been introduced. Note that we removed the terms that include the bare Green's function, $G_{0}(\bold{r},\bold{r}')$, since they pertain to identically divergenceless tensor and do not contribute to mechanical forces.

We can determine the average over the Gaussian measure using the expression
\begin{equation}
\left<(\cdot)\right> _{0} = \int \frac{\mathcal{D}\eta}{C_1} e^{-\frac{1}{2}(\eta G^{-1}\eta)} (\cdot),
\end{equation}
where $C_1$ is defined as
\begin{equation}
C_1=\int \mathcal{D}\eta e^{-\frac{1}{2}(\eta G^{-1}\eta)}.
\end{equation}
By applying Wick's theorem~\cite{zinn2002quantum_} from equation (\ref{16}), we can obtain the expression (\ref{1.18}) as previously stated. Eq. (\ref{8_2}) being the main result of this paper determines the fluctuation contribution to the total stress tensor rising from the thermal fluctuations of the local electrostatic potential and field near their mean-field configuration. It determines the contribution of fluctuations to the total stress tensor, which arises from the thermal fluctuations of the local electrostatic potential and field near their mean-field configuration. In other words, this tensor describes the effect of electrostatic correlation on local stresses in the inhomogeneous Coulomb fluid. It is called the {\sl correlation stress tensor}.

In order to calculate the macroscopic force acting on the dielectric or conducting body immersed in Coulomb fluid, it is necessary to solve the EL equation (\ref{1.11}) for $\varphi(\bold{r})$ with appropriate boundary conditions and equation for the Green's function (\ref{1.16}) and then calculate the following surface integral over the body surface \cite{budkov2022modified,budkov2023macroscopic}
\begin{equation}
F_{i}=\oint\limits_{\mathcal{A}}\sigma_{ik}n_{k} d\mathcal{A},
\end{equation}
where $n_{k}$ is the external normal and $d\mathcal{A}$ is the elementary area.

\section{Stress tensor. General covariant approach}
Now, we will discuss an alternative, more sophisticated approach to obtaining the total stress tensor from the GTP derived above eq.~(\ref{GTP}). This approach is based on the general covariant methodology presented in our recent work~\cite{brandyshev2023noether}. As it was already mentioned in Introduction, in the present case, this approach does not have any advantages relative to the one based on Noether's theorem. However, we would like to consider it below from a pedagogical standpoint.

In this approach, the stress tensor can be obtained using the following expression
\begin{equation}
\label{sigma_deriv_metric}
\sigma_{ik}=\frac{2}{\sqrt{g}}\frac{\delta \Omega}{\delta g_{ik}}\bigg{|}_{g_{ik}=\delta_{ik}},
\end{equation}
where $g_{ij}$ is the metric tensor, and $g=\det{g_{ij}}$ -- its determinant and $\Omega$ is the GTP (\ref{GTP}) obtained above within the functional Legendre transformation approach.

In order to apply eq. (\ref{sigma_deriv_metric}), we have to express the GTP in general covariant form. Thus, we have 
\begin{equation}\label{36}
\Omega[i\varphi]=\mathcal{H}[i\varphi]+\mathcal{H}^{\prime}[\varphi],\quad \mathcal{H}^{\prime}[\varphi]=\frac{\Omega_1[\varphi]}{2\beta},
\end{equation}
where
\begin{equation}
\mathcal{H}=\int d\bold{r} \sqrt{g} 
\bigg[-\frac{\varepsilon}{2}g^{ij}\partial_{i}\varphi\partial_{j}\varphi-P(\{\bar{\mu}_{\alpha}\})\bigg],
\end{equation}
is the "mean-field" functional with the aforementioned shifted chemical potentials $\bar{\mu}_{\alpha}=\mu_{\alpha}-q_{\alpha}\varphi$.
Note that we have implied the summation over repeated coordinate indices. We also assumed here for simplicity $u_{\alpha}=0$.

The one-loop correction is
\begin{equation}\label{1}
\mathcal{H}^{\prime}[\varphi]=\frac{k_BT}{2}tr\ln\bigg(\frac{-\Delta+U(\varphi)}{-\Delta}\bigg),
\end{equation}
where Laplacian can be written in general covariant form~\cite{landau1971classical,weinberg1972gravitation}
\begin{equation}\label{}
\Delta f=\frac{1}{\sqrt{g}}\partial_{i}\bigg(\sqrt{g}g^{ij}\partial_{j}f\bigg).
\end{equation}
General coordinate transformations lead to
\begin{equation}\label{}
d\bold{r}\rightarrow Jd\bold{r},\quad \sqrt{g}\rightarrow J^{-1}\sqrt{g},
\end{equation}
where $J$ is the Jacobian determinant. Thus, the invariant delta-function can be defined by 
\begin{equation}\label{18}
\int d\bold{r}\sqrt{g(\bold{r})}\delta(\bold{r})=1.
\end{equation}
Let us consider an infinitesimal transformation of the metric tensor
\begin{equation}\label{17}
g_{ij}(\bold{r})\rightarrow g_{ij}(\bold{r})+\delta g_{ij}(\bold{r}).
\end{equation}
Then, we have~\cite{landau1971classical}
\begin{equation}\label{5}
\delta(\sqrt{g})=\frac{1}{2}\sqrt{g}g^{ij}\delta g_{ij},\quad \delta g^{ij}=-g^{im}g^{jn}\delta g_{mn}.
\end{equation}

The zeroth order term is 
\begin{equation}
\sigma^{(0)}_{ik}=\frac{2}{\sqrt{g}}\frac{\delta\mathcal{H}[i\varphi]}{\delta g_{ik}}\bigg|_{g_{ik}=\delta_{ik}},
\end{equation}
that yields the same functional form that realized in the mean-field approximation, i.e.
\begin{equation}
\sigma^{(0)}_{ik}=\varepsilon\partial_i\varphi\partial_k\varphi-
\frac{\varepsilon}{2}\delta_{ik}\partial_j\varphi\partial_j\varphi-P\delta_{ik}.
\end{equation}
However, it should not be confused with the mean-field approximation~(\ref{sigma_MF}), since the electrostatic potential $\varphi$ satisfies the equation of EL (\ref{1.11}) taking into account the electrostatic correlations. It can be shown that the divergence of this tensor is 
\begin{equation}\label{33}
\partial_i\sigma^{(0)}_{ik}=\bigg(-\frac{\partial P(\varphi)}{\partial\varphi}+\varepsilon\Delta\varphi\bigg)\partial_k\varphi.
\end{equation}

Expanding $\sigma^{(0)}_{ik}$ in series on $\chi$ and holding first order terms in $k_{B}T$, and expanding the chemical potentials $\mu_{\alpha}=\mu_{\alpha}^{(0)}+\mu_{\alpha}^{(1)}$ up to the first order in $k_{B}T$, we get
\begin{equation}
\label{sigma0}
\sigma^{(0)}_{ik}=\sigma^{(MF)}_{ik}-\delta_{ik}\sum\limits_{\lambda}\mu_{\lambda}^{(1)}\bar{c}_{\lambda}+\frac{\varepsilon}{2\beta}\bigg(\partial_i\chi\partial_k\psi+\partial_i\psi\partial_k\chi-
\delta_{ik}\partial_j\chi\partial_j\psi\bigg)-\delta_{ik} \frac{\chi}{2\beta}\frac{\partial P}{\partial \psi}.
\end{equation}
The third and forth  terms in the right hand side of eq. (\ref{sigma0}) determine a contribution to the total stress tensor arising from the mismatch between the electrostatic potential of the mean-field, $\psi$, and its expectation value, $\varphi$. 

Thus, the stress tensor is
\begin{equation}
\sigma_{ik}=\sigma^{(0)}_{ik}+\sigma^{(1)}_{ik},
\end{equation}
where the first-order term can be obtained in the same way
\begin{equation}\label{32}
\sigma^{(1)}_{ik}=\frac{2}{\sqrt{g}}\frac{\delta\mathcal{H}^{\prime}}{\delta g_{ik}}\bigg|_{g_{ik}=\delta_{ik}}.
\end{equation}
In order to perform the calculation in eq. (\ref{32}), we have to introduce a general covariant definition of the trace, i.e.
\begin{equation}\label{2}
tr(A)=\int d\bold{r} \sqrt{g(\bold{r})}A(\bold{r},\bold{r}),
\end{equation}
For composition of two integral operators,
\begin{equation}\label{}
C=AB,
\end{equation}
we can introduce the rule
\begin{equation}
\label{3}
C(\bold{r},\bold{r}')=\int d\bold{r}'' \sqrt{g(\bold{r}'')}A(\bold{r},\bold{r}'')B(\bold{r}'',\bold{r}').
\end{equation}
Action of the operator $A$ on a function $f(\bold{r})$ is determined by
\begin{equation}\label{4}
Af(\bold{r})=\int d\bold{r}' \sqrt{g(\bold{r}')} A(\bold{r},\bold{r}')f(\bold{r}').
\end{equation}
The trace variation is
\begin{equation}
\label{}
\delta tr(A)=\int d\bold{r} \sqrt{g(\bold{r})}\delta A(\bold{r},\bold{r})+\frac{1}{2}\int d\bold{r} \sqrt{g(\bold{r})}g^{ij}(\bold{r})\delta g_{ij}(\bold{r}) A(\bold{r},\bold{r}),
\end{equation}
which can be rewritten as
\begin{equation}\label{8}
\delta tr(A)=tr(\bar{\delta} A),
\end{equation}
where we have introduced the infinitesimal operator $\bar{\delta} A$ that has the kernel determined by the identity
\begin{equation}\label{24}
\bar{\delta} A(\bold{r},\bold{r}')=\delta A(\bold{r},\bold{r}')+\frac{1}{2} A(\bold{r},\bold{r}')g^{ij}(\bold{r}')\delta g_{ij}(\bold{r}').
\end{equation}
Thus, we have
\begin{equation}\label{}
\delta tr(A)=\int d\bold{r} \sqrt{g(\bold{r})}\bar{\delta} A(\bold{r},\bold{r}).
\end{equation}
Using eq. (\ref{5}), we can show that the operator $\Delta$ is transformed as
\begin{equation}\label{}
\Delta\rightarrow \Delta+\Delta',
\end{equation}
where
\begin{equation}\label{12_1}
\begin{aligned}
\Delta'f=\frac{1}{2\sqrt{g}}\partial_{i}\bigg(\sqrt{g}g^{mn}\delta g_{mn}g^{ij}\partial_{j}f\bigg)-
\frac{1}{2\sqrt{g}}g^{mn}\delta g_{mn}\partial_{i}\bigg(\sqrt{g}g^{ij}\partial_{j}f\bigg)\\
-\frac{1}{\sqrt{g}}\partial_{i}\bigg(\sqrt{g}\delta g_{mn}g^{im}g^{jn}\partial_{j}f\bigg),
\end{aligned}
\end{equation}
Functional (\ref{1}) can be rewritten as
\begin{equation}\label{}
\mathcal{H}^{\prime}=\frac{k_{B}T}{2}tr\ln\bigg(I+\frac{\varepsilon G_{0}U}{k_{B}T}\bigg),
\end{equation}
where operator $G_0$ is determined by equation
\begin{equation}\label{6_1}
-\frac{\varepsilon}{k_{B}T}\Delta G_{0}=I,
\end{equation}
where $I$ is the unity operator. Varying both sides of eq. (\ref{6_1}) on metric tensor, we arrive at (see Appendix A)
\begin{equation}\label{25}
-\frac{\varepsilon}{k_{B}T}\bigg(\Delta' G_{0}+\Delta\bar{\delta} G_{0}\bigg)=0.
\end{equation}
that in turn yields  
\begin{equation}\label{}
\bar{\delta} G_{0}=\frac{\varepsilon}{k_{B}T}G_{0}\Delta'G_{0}.
\end{equation} 
Thus, the functional variation of the fluctuation correction is
\begin{equation}\label{}
\delta\mathcal{H}^{\prime}=\frac{\varepsilon}{2}tr\bigg[\bigg(I+\frac{\varepsilon G_{0}U}{k_{B}T}\bigg)^{-1}
\bar{\delta} G_{0}U\bigg].
\end{equation}
Therefore, we obtain
\begin{equation}\label{}
\delta\mathcal{H}^{\prime}=\frac{\varepsilon}{2}tr\bigg[\bigg(I+\frac{\varepsilon G_{0}U}{k_{B}T}\bigg)^{-1}
\frac{\varepsilon}{k_{B}T}G_{0}\Delta'G_{0}U\bigg].
\end{equation}
Let us introduce the operator $G$, determined by the identity (\ref{1.16}) that can be rewritten as follows
\begin{equation}\label{7}
\frac{\varepsilon}{k_{B}T}\bigg(-\Delta+U\bigg)G=I.
\end{equation}
Then using (\ref{6_1}) and (\ref{7}), we can get
\begin{equation}\label{9}
\bigg(I+\frac{\varepsilon G_{0}U}{k_{B}T}\bigg)^{-1}=GG_{0}^{-1}.
\end{equation}
Thus, the functional variation is
\begin{equation}\label{10}
\delta\mathcal{H}^{\prime}=\frac{\varepsilon^{2}}{2k_{B}T}tr(G\Delta'G_0U)
\end{equation}
Using eq. (\ref{9}) and taking into account (\ref{6_1}), we arrive at
\begin{equation}\label{}
\frac{\varepsilon G_{0}U}{k_{B}T}=G_{0}G^{-1}-I.
\end{equation}
Thus, by using the invariance of the trace with respect to cyclic permutations of operators in eq. (\ref{10}), we can obtain the following
\begin{equation}\label{}
\begin{aligned}
\delta\mathcal{H}^{\prime}=-\frac{\varepsilon}{2}tr(\Delta'\bar{G}),
\end{aligned}
\end{equation}
where
\begin{equation}
\bar{G}=G-G_0.
\end{equation}
Expression (\ref{12_1}) can be rewritten as
\begin{equation}\label{}
\begin{aligned}
\Delta'f=Df-
\frac{1}{2}g^{mn}\delta g_{mn}\Delta f,
\end{aligned}
\end{equation}
where we have introduced the following auxiliary differential operator
\begin{equation}\label{}
\begin{aligned}
Df=\frac{1}{2\sqrt{g}}\partial_{i}\bigg(\sqrt{g}g^{mn}\delta g_{mn}g^{ij}\partial_{j}f\bigg)
-\frac{1}{\sqrt{g}}\partial_{i}\bigg(\sqrt{g}\delta g_{mn}g^{im}g^{jn}\partial_{j}f\bigg).
\end{aligned}
\end{equation}
Then, we have
\begin{equation}
\label{14}
\begin{aligned}
\delta\mathcal{H}^{\prime}=-\frac{\varepsilon}{2}tr(D\bar{G})+\frac{\varepsilon}{4}tr\bigg(g^{mn}\delta g_{mn}\Delta \bar{G}\bigg),
\end{aligned}
\end{equation}
where
\begin{equation}\label{}
\begin{aligned}
tr(DG)=\int d\bold{r} \int d\bold{r}' \sqrt{g(\bold{r}')}G(\bold{r}',\bold{r})\sqrt{g(\bold{r})}D_\bold{r}\delta(\bold{r}-\bold{r}')
\end{aligned}
\end{equation}
and for brevity we use the notation $G(\bold{r},\bold{r}')=G(\bold{r},\bold{r}'|\varphi)$.

Integrating it by parts with account of $\delta g_{mn}=0$ on the boundary of integration, we get
\begin{equation}\label{}
tr(DG)=-\int d\bold{r}' \int d\bold{r} \sqrt{g(\bold{r}')}\partial_{i}G(\bold{r}',\bold{r})\sqrt{g(\bold{r})}\delta g_{mn}(\bold{r})\partial_{j}\delta(\bold{r}-\bold{r}')\gamma^{ijmn}(\bold{r}),
\end{equation}
where we have introduced the following tensor function
\begin{equation}\label{}
\gamma^{ijmn}(\bold{r})=\frac{1}{2}g^{mn}(\bold{r})g^{ij}(\bold{r})
-g^{im}(\bold{r})g^{jn}(\bold{r}).
\end{equation}
Thus, we have
\begin{equation}\label{}
\begin{aligned}
\int d\bold{r}' \sqrt{g(\bold{r}')}\partial_{i}G(\bold{r}',\bold{r})\partial_{j}\delta(\bold{r}-\bold{r}')=
\lim_{\bold{r}'\rightarrow \bold{r}}\partial_{j}\int d\bold{r}'' \sqrt{g(\bold{r}'')}\partial_{i}'G(\bold{r}'',\bold{r}')\delta(\bold{r}-\bold{r}'')
\end{aligned}
\end{equation}
so that 
\begin{equation}\label{}
\begin{aligned}
\int d\bold{r}' \sqrt{g(\bold{r}')}\partial_{i}G(\bold{r}',\bold{r})\partial_{j}\delta(\bold{r}-\bold{r}')=
\lim_{\bold{r}'\rightarrow \bold{r}}\partial_{j}\partial_{i}'G(\bold{r},\bold{r}').
\end{aligned}
\end{equation}
Therefore, we obtain
\begin{equation}\label{}
tr(DG)=-\int d\bold{r} \sqrt{g(\bold{r})}\delta g_{mn}(\bold{r})\gamma^{ijmn}(\bold{r})\lim_{\bold{r}'\rightarrow \bold{r}}\partial_{j}\partial_{i}'G(\bold{r},\bold{r}').
\end{equation}
The second term in eq. (\ref{14}) is
\begin{equation}\label{}
tr\bigg(g^{mn}\delta g_{mn}\Delta G\bigg)=
\int d\bold{r}'g^{mn}(\bold{r}')\delta g_{mn}(\bold{r}')\sqrt{g(\bold{r}')}\int d\bold{r}\sqrt{g(\bold{r})}G(\bold{r},\bold{r}')\Delta_{\bold{r}'}\delta(\bold{r}'-\bold{r}).
\end{equation}
Using the expression
\begin{equation}\label{}
\int d\bold{r}\sqrt{g(\bold{r})}G(\bold{r},\bold{r}')\Delta_{\bold{r}'}\delta(\bold{r}'-\bold{r})=\lim_{\bold{r}''\rightarrow \bold{r}'}\Delta_{\bold{r}'}\int d\bold{r}\delta(\bold{r}'-\bold{r})\sqrt{g(\bold{r})}G(\bold{r},\bold{r}''),
\end{equation}
we can write
\begin{equation}\label{}
tr\bigg(g^{mn}\delta g_{mn}\Delta G\bigg)=\\
\int d\bold{r}'g^{mn}(\bold{r}')\delta g_{mn}(\bold{r}')\sqrt{g(\bold{r}')}\lim_{\bold{r}''\rightarrow \bold{r}'}\bigg(\Delta_{\bold{r}'}G(\bold{r}',\bold{r}'')\bigg)
\end{equation}
Then, using eq. (\ref{32}), we can get
\begin{equation}\label{35}
\sigma^{(1)}_{ij}(\bold{r})=\frac{\varepsilon}{2}\lim_{\bold{r}'\rightarrow \bold{r}}\bigg(\delta_{ij}\partial_k\partial_k \bar{G}(\bold{r},\bold{r}')+\delta_{ij}\partial_k\partial_k' \bar{G}(\bold{r},\bold{r}')-\partial_i\partial_j' \bar{G}(\bold{r},\bold{r}')-\partial_j\partial_i' \bar{G}(\bold{r},\bold{r}')\bigg).
\end{equation}
The latter can be expressed in more compact form
\begin{equation}\label{}
\begin{aligned}
\sigma^{(1)}_{ij}(\bold{r})=\frac{\varepsilon}{2}\lim_{\bold{r}'\rightarrow \bold{r}}\hat{D}_{ij}\bar{G}(\bold{r},\bold{r}'),
\end{aligned}
\end{equation}
where the following differential operator
\begin{equation}
\hat{D}_{ij}=\delta_{ij}\partial_k\partial_k +\delta_{ij}\partial_k\partial_k'\\
-\partial_i\partial_j'-\partial_j\partial_i'
\end{equation}
has been introduced.

The divergence of this tensor is
\begin{equation}\label{}
\begin{aligned}
\partial_{i}\sigma^{(1)}_{ij}(\bold{r})=\frac{\varepsilon}{2}\lim_{\bold{r}'\rightarrow \bold{r}}
\bigg[\bigg(\partial_{i}+\partial_{i}'\bigg)\bigg(\delta_{ij}\partial_k\partial_k \bar{G}(\bold{r},\bold{r}')+\delta_{ij}\partial_k\partial_k'\bar{G}(\bold{r},\bold{r}')\\
-\partial_i\partial_j'\bar{G}(\bold{r},\bold{r}')-\partial_j\partial_i'\bar{G}(\bold{r},\bold{r}')\bigg)\bigg].
\end{aligned}
\end{equation}
Therefore, after some algebra, we can get
\begin{equation}\label{a}
\begin{aligned}
\partial_{i}\sigma^{(1)}_{ik}(\bold{r})=\frac{\varepsilon}{2}\lim_{\bold{r}'\rightarrow \bold{r}}
\bigg[\partial_{k}\partial_{i}\partial_{i}\bar{G}(\bold{r},\bold{r}')-\partial_{i}'\partial_{i}'\partial_{k}\bar{G}(\bold{r},\bold{r}')\bigg],
\end{aligned}
\end{equation}
and using eqs. (\ref{6_1}) and (\ref{7}), we rewrite eq. (\ref{a}) in the form
\begin{equation}\label{34}
\begin{aligned}
\partial_{i}\sigma^{(1)}_{ik}(\bold{r})=\frac{\varepsilon}{2}G(\bold{r},\bold{r})\partial_{k}U(\varphi),
\end{aligned}
\end{equation}
where
\begin{equation}\label{}
\partial_{k}U(\varphi)=\frac{\partial U(\varphi)}{\partial\varphi}\partial_{k}\varphi.
\end{equation}
Thus, using eqs. (\ref{33}) and (\ref{34}), we obtain
\begin{equation}\label{37}
\partial_{i}\sigma_{ik}(\bold{r})=\frac{\delta\Omega[\varphi]}{\delta\varphi(\bold{r})}\partial_{k}\varphi(\bold{r}).
\end{equation}
where
\begin{equation}\label{}
\frac{\delta\Omega[\varphi]}{\delta\varphi(\bold{r})}=-\frac{\partial P(\varphi)}{\partial\varphi}+\varepsilon\Delta\varphi+\frac{\varepsilon}{2}\frac{\partial U(\varphi)}{\partial\varphi(\bold{r})}G(\bold{r},\bold{r}).
\end{equation}
Thereby, the local mechanical equilibrium condition~\cite{budkov2022modified,hermann2022noether,sprik2021chemomechanical},
\begin{equation}
\partial_{i}\sigma_{ik}=0,
\end{equation}
is satisfied if $\varphi$ is a solution of EL equation (\ref{1.11}). Thus, we have demonstrated the exactness of eq. (\ref{37}) with respect to the terms of any order in $1/\beta$, for the functional (\ref{36}). However, as the functional (\ref{36}) was only calculated with precision up to the first-order terms, solving equation (\ref{1.11}) with higher-order accuracy would not be meaningful.\par

Further, we can show that the tensor (\ref{35}) can be rewritten in the form
\begin{equation}\label{fluct_corr_2}
\begin{aligned}
\sigma^{(1)}_{ij}(\bold{r})=\sigma^{(cor)}_{ij}(\bold{r})=\frac{\varepsilon}{2}\bigg(U(\psi)Q(\bold{r})
+\mathcal{D}_{kk}(\bold{r})\bigg)\delta_{ij}-\varepsilon \mathcal{D}_{ij}(\bold{r}),
\end{aligned}
\end{equation}
where we took into account terms of first order on $1/\beta$ and introduce the following short-hand notations
\begin{equation}
Q(\bold{r})=\bigg\langle \eta^2(\bold{r}) \bigg\rangle =\lim_{\bold{r}'\rightarrow \bold{r}}G(\bold{r},\bold{r}'|\psi),
\end{equation}
\begin{equation}
\mathcal{D}_{ij}(\bold{r})= \lim_{\bold{r}'\rightarrow \bold{r}}
\partial_i\partial_j' G(\bold{r},\bold{r}'|\psi).
\end{equation}
where the expectation values are calculated in the Gaussian approximation derived from the expansion near the mean-field configuration.

Thus, we arrive at the same expression for the correlation stress tensor (\ref{8_2}), which has been obtained above within the Noether theorem-based approach. Note that, as was done above, we removed the terms that include the bare Green function, $G_{0}(\bold{r},\bold{r}')$, since they pertain to identically divergenceless tensors and do not contribute to mechanical forces.

We would like to point out that the formulation of our theory bears formal similarities to that of the theory of van der Waals forces by E.M. Lifshitz \cite{lifshitz2013statistical,dzyaloshinskii1961general}. Nevertheless, our theory deals with pure classical electrostatic fluctuations, as opposed to the electromagnetic quantum fluctuations presented in the Lifshitz theory.

\section{1D case}
In this section, we would like to specify the correlation stress tensor obtained above for a practically important one-dimensional case, i.e. when the Coulomb fluid is confined in a slit-like pore or in close proximity to a flat electrified surface. Using the cylindrical coordinates, we can write 
\begin{equation}\label{}
G(\bold{r},\bold{r}')=G(\bm{\rho}-\bm{\rho}'|z,z'),
\end{equation}
where $\bm{\rho}$ is the two-dimensional vector lying in the plane of the pore and the $z$-axis is perpendicular to the pore plane. Considering the Fourier transform
\begin{equation}\label{}
G(\bm{\rho}-\bm{\rho}'|z,z')=\int \frac{d^2\bold{q}}{(2\pi)^2}e^{-i\bold{q}(\bm{\rho}-\bm{\rho}')}G(q|z,z'),
\end{equation}
where $G(q|z,z')$ is the even function of $\bold{q}$ depending only on the vector modulus $q=|\bold{q}|$. This system is one-dimensional in the sense that the functions $\psi$ and $\mathcal{D}_{ij}$ depend only on $z$. Let us first calculate the cross elements
\begin{equation}\label{}
\mathcal{D}_{xy}(z)=\lim_{z'\rightarrow z}\int \frac{d^2\bold{q}}{(2\pi)^2}q_xq_yG(q|z,z'),
\end{equation}
\begin{equation}\label{}
\mathcal{D}_{xz}(z)=\lim_{z'\rightarrow z}\int \frac{d^2\bold{q}}{(2\pi)^2}iq_x\partial_zG(q|z,z'),
\end{equation}
\begin{equation}\label{}
\mathcal{D}_{yz}(z)=\lim_{z'\rightarrow z}\int \frac{d^2\bold{q}}{(2\pi)^2}iq_y\partial_zG(q|z,z'),
\end{equation}
It is obvious that these elements are zero, since the integral of an odd function is zero
\begin{equation}\label{}
\mathcal{D}_{xy}(z)=\mathcal{D}_{xz}(z)=\mathcal{D}_{yz}(z)=0.
\end{equation}
Diagonal elements are
\begin{equation}\label{}
\mathcal{D}_{xx}(z)=\int \frac{d^2\bold{q}}{(2\pi)^2}q_x^2Q(q,z),~\mathcal{D}_{yy}(z)=\int \frac{d^2\bold{q}}{(2\pi)^2}q_y^2Q(q,z),
\end{equation}
\begin{equation}\label{}
\mathcal{D}_{zz}(z)=\int \frac{d^2\bold{q}}{(2\pi)^2}\mathcal{D}_{zz}(q,z),
\end{equation}
where
\begin{equation}
\mathcal{D}_{zz}(q,z)=\lim_{z'\rightarrow z}\partial_z\partial_{z'}G(q|z,z'),
\end{equation}
\begin{equation}
Q(q,z)=\lim_{z'\rightarrow z}G(q|z,z').
\end{equation}
Thus the trace is
\begin{equation}
\mathcal{D}_{kk}(z)=\int \frac{d^2\bold{q}}{(2\pi)^2}\bigg(q^2Q(q,z)+\mathcal{D}_{zz}(q,z)\bigg).
\end{equation}
The Fourier image of the Green's function can be found from the equation
\begin{equation}\label{}
\varepsilon \beta \bigg(-\partial^2_z+q^2+\varkappa^2(z)\bigg)G(q|z,z')=\delta(z-z').
\end{equation}
The correlation stress tensor elements are defined by the following expressions
\begin{equation}\label{sigma1}
\begin{aligned}
\sigma^{(cor)}_{ij}(z)=\int \frac{d^2\bold{q}}{(2\pi)^2}\sigma^{(cor)}_{ij}(q,z),
\end{aligned}
\end{equation}
\begin{equation}\label{sigma2}
\begin{aligned}
\sigma^{(cor)}_{xx}(q,z)=\frac{\varepsilon}{2}\bigg(Q(q,z)\varkappa^2(z)
+\mathcal{D}_{zz}(q,z)\bigg),
\end{aligned}
\end{equation}
\begin{equation}\label{sigma3}
\begin{aligned}
\sigma^{(cor)}_{yy}(q,z)=\sigma^{(cor)}_{xx}(q,z),
\end{aligned}
\end{equation}
\begin{equation}\label{sigma4}
\begin{aligned}
\sigma^{(cor)}_{zz}(q,z)=\frac{\varepsilon}{2}\bigg(Q(q,z)\left(q^2+\varkappa^2(z)\right)
-\mathcal{D}_{zz}(q,z)\bigg).
\end{aligned}
\end{equation}

Note that the formulas derived formulas (\ref{sigma1}-\ref{sigma4}) in case of Coulomb gas (point-like ions) for very wide pores yield expressions for the stresses in the bulk phase fluid, that is, Debye-H{\"u}ckel expression (see Appendix B).

\section{Concluding remarks}
In conclusion, we have developed a first-principles statistical field theory for analyzing the macroscopic forces in spatially inhomogeneous Coulomb fluids. Additionally, we have generalized Noether's first theorem to account for a fluctuating order parameter and have derived the total stress tensor using this formulation. By employing this methodology, we have obtained the stress tensor for the Coulomb fluid, which includes both the previously derived mean-field stress tensor and fluctuation corrections at the one-loop correction level. These fluctuation corrections incorporate the thermal fluctuations of the local electrostatic potential and field around the mean-field configuration. The correlation stress tensor, derived from these fluctuation corrections, reflects how electrostatic correlation influences the local stresses in a non-uniform Coulomb fluid. Furthermore, by combining the previously formulated general covariant approach with the functional Legendre transform method, we have reproduced the same result for the stress tensor.

The developed formalism is interesting for modeling not only inhomogeneous Coulomb fluids but also bulk fluids. In particular, it is intriguing to analyze how this formalism can predict the screening length for concentrated electrolyte solutions, considering the excluded volume of the ions, in comparison with other methods of stastistical physics~\cite{attard1993asymptotic,kjellander2020multiple}. This issue might be the subject of future publications.

Finally, we would like to note that proposed approaches is applicable not only to the Coulomb fluids, but also to nonionic simple or complex fluids (including quantum ones) for which we know the field-theoretic Hamiltonian as a functional of the appropriate fluctuating order parameters.

{\bf Acknowledgements.}  
This work is an output of a research project implemented as part of the Basic Research Program at the National Research University Higher School of Economics (HSE University).

\appendix
\section{}
Let us introduce the function $\tilde{\delta}(\bold{r}-\bold{r}')$ determined by expression
\begin{equation}\label{19}
\int d\bold{r}\sqrt{\tilde{g}(\bold{ r})}\tilde{\delta}(\bold{r}-\bold{r}')f(\bold{r})=f(\bold{r}').
\end{equation}
where we consider the metric variation as the type function variation (it is not variation under coordinate transformations)
\begin{equation}\label{20}
\tilde{g}_{ij}(\bold{r})= g_{ij}(\bold{r})+\delta g_{ij}(\bold{r}),
\end{equation}
and then we can define the variation
\begin{equation}\label{}
\delta'(\bold{r}-\bold{r}')=\tilde{\delta}(\bold{r}-\bold{r}')-\delta(\bold{r}-\bold{r}').
\end{equation}
Then, using eq. (\ref{18}), we can show that the equality (\ref{19}) is fulfilled if the variation of delta-function is determined by the following identity
\begin{equation}\label{23}
\delta'(\bold{r}-\bold{r}')=-\frac{1}{2}g^{ij}(\bold{r}')\delta g_{ij}(\bold{r}')\delta(\bold{r}-\bold{r}').
\end{equation}
Let us note that using known delta-function properties we can write
\begin{equation}
g^{ij}(\bold{r}')\delta g_{ij}(\bold{r}')\delta(\bold{r}-\bold{r}')=g^{ij}(\bold{r})\delta g_{ij}(\bold{r})\delta(\bold{r}-\bold{r}').
\end{equation}
It is obviously that in this case 
\begin{equation}
\delta\bigg(\sqrt{g(\bold{r}')}\delta(\bold{r}-\bold{r}')\bigg)=0.
\end{equation}
Under transformations (\ref{20}) the operator $G_0$ is infinitesimally transformed as follows
\begin{equation}\label{}
\tilde{G_{0}}=G_{0}+\delta G_{0}.
\end{equation}
where the transformed operator $\tilde{G_{0}}$ is determined by the equation
\begin{equation}\label{22}
-\frac{\varepsilon}{k_{B}T}\bigg(\Delta_{\bold{r}}+\Delta'_{\bold{r}}\bigg)\tilde{G_{0}}(\bold{r},\bold{r}')=\tilde{\delta}(\bold{r}-\bold{r}').
\end{equation}
The equation (\ref{6}) can be written in the form
\begin{equation}\label{21}
-\frac{\varepsilon}{k_{B}T}\Delta_{\bold{r}}G_{0}(\bold{r},\bold{r}')=\delta(\bold{r}-\bold{r}').
\end{equation}
Then subtracting (\ref{21}) from expression (\ref{22}) with account of eq. (\ref{23}) and holding only first-order terms, we can get
\begin{equation}\label{}
-\frac{\varepsilon}{k_{B}T}\bigg(\Delta'_{\bold{r}}G_{0}(\bold{r},\bold{r}')+\Delta_{\bold{r}}\delta G_{0}(\bold{r},\bold{r}')\bigg)=-\frac{1}{2}g^{ij}(\bold{r}')\delta g_{ij}(\bold{r}')\delta(\bold{r}-\bold{r}').
\end{equation}
Using eq. (\ref{21}) again, we can rewrite this equality in the form
\begin{equation}\label{26}
-\frac{\varepsilon}{k_{B}T}\bigg(\Delta'_{\bold{r}} G_{0}(\bold{r},\bold{r}')+\Delta_{\bold{r}}\delta G_{0}(\bold{r},\bold{r}')+\frac{1}{2}g^{ij}(\bold{r}')\delta g_{ij}(\bold{r}')\Delta_{\bold{r}} G_{0}(\bold{r},\bold{r}')\bigg)=0.
\end{equation}
Using definition (\ref{24}), eq. (\ref{26}) can be reduced to eq. (\ref{25}).\par
Let us consider the following expressed diffeomorphic transformations
\begin{equation}\label{28}
\begin{aligned}
\tilde{x}^{k}=x^{k}+\xi^{k}(\bold{r}),\\
\tilde{g}_{ij}(\tilde{\bold{r}})= g_{ij}(\bold{r})+\tilde{\delta} g_{ij}(\bold{r}),
\end{aligned}
\end{equation}
where $\xi^{k}$ and $\tilde{\delta} g_{ij}$ are generally independent transformation parameters, $\tilde{x}^{k}$ and $x^{k}$ are $k$th components of vectors $\tilde{\bold{r}}$ and $\bold{r}$, respectively. Then, using the identity
\begin{equation}\label{}
\int d\tilde{\bold{r}}_1\sqrt{\tilde{g}(\tilde{\bold{r}}_1)}\tilde{\delta}(\tilde{\bold{r}}_1-\tilde{\bold{r}}_2)f(\tilde{\bold{r}}_1)=f(\tilde{\bold{r}}_2).
\end{equation}
we can show that the delta-function is transformed under (\ref{28}), in accordance with the expression
\begin{equation}\label{}
\tilde{\delta}(\tilde{\bold{r}}_1-\tilde{\bold{r}}_2)=\bigg(1-\partial_{k} \xi^{k}(\bold{r}_1)-
\frac{1}{2}g^{ij}(\bold{r}_1)\tilde{\delta} g_{ij}(\bold{r}_1)\bigg)\delta(\bold{r}_1-\bold{r}_2).
\end{equation}
and if we put that
\begin{equation}\label{30}
\tilde{\delta}g^{ik}(\bold{r})=\xi^{i,k}(\bold{r})+\xi^{k,i}(\bold{r}),
\end{equation}
where
\begin{equation}
 \xi^{i,k}(\bold{r})=g^{kj}\partial_{j} \xi^{i}(\bold{r}),   
\end{equation}
then in this special case
\begin{equation}\label{}
\tilde{\delta}(\tilde{\bold{r}}_1-\tilde{\bold{r}}_2)=\delta(\bold{r}_1-\bold{r}_2).
\end{equation}
The delta function thus defined is invariant under a subgroup of transformations (\ref{28}) that correspond to eq. (\ref{30}) (under standard diffeomorphisms). Our approach requires us to consider another subgroup of (\ref{28}) that corresponds to the condition that $\xi^k=0$, and where $\tilde{\delta} g_{ij}$ can be an arbitrary infinitesimal function. The delta function is transformed under this subgroup according to the law (\ref{23}).

\section{}
By analyzing the asymptotic behavior of the normal stresses for extremely large pore thicknesses, $H$, it is possible to determine whether the formulated formalism can accurately describe the stresses in the bulk Coulomb fluids. Specifically, the analysis would focus on the behavior of the normal stresses at $H\to \infty$. In this case, $\varkappa(z)=\kappa=$const (inverse screening length of bulk Coulomb fluid~\cite{maggs2016general}) and 
\begin{equation}
G(q|z,z^{\prime})\simeq \frac{\exp\left(-\kappa_q|z-z^{\prime}|\right)}{2\beta\varepsilon\kappa_q},
\end{equation}
so that we have
\begin{equation}
Q(q,z)\simeq\frac{1}{2\beta\varepsilon \kappa_q},~\mathcal{D}_{zz}(q,z)\simeq -\frac{\kappa_q}{2\beta\varepsilon},
\end{equation}
where $\kappa_q=\sqrt{\kappa^2+q^2}$. Thus, we have for the normal correlation stress
\begin{equation}
\sigma_{zz}^{(cor)}=\frac{k_{B}T}{2}\int\frac{d^2\bold{q}}{(2\pi)^2}\left(\kappa_{q}-q-\frac{\kappa^2}{2q}\right)=-\frac{k_{B}T\kappa^3}{12\pi},
\end{equation}
where we have subtracted from the final result the infinite value. To obtain the total normal stress at $H\to \infty$, we have to calculate (see eq. (\ref{sigma0}))
\begin{equation}
\sigma_{zz}^{(0)}=-k_{B}T\sum\limits_{\alpha}c_{\alpha}-\sum\limits_{\alpha}c_{\alpha}\mu_{\alpha}^{(1)},
\end{equation}
where $c_{\alpha}$ are the bulk ionic concentrations. This contribution could be easily obtained for the case, when the reference system is the mixture of the ideal gases, for which  
\begin{equation}
P(\{\bar{\mu}_{\alpha}\})=k_{B}T\sum\limits_{\alpha}\Lambda_{\alpha}^{-3}e^{\beta\bar{\mu}_{\alpha}},
\end{equation}
where $\Lambda_{\alpha}$ are the de Broglie thermal wavelengths. In this case, we have
\begin{align}
\bar{c}_{\alpha}(\bold{r})=\frac{\delta \Omega}{\delta u_{\alpha}(\bold{r})}=\Lambda_{\alpha}^{-3}e^{\beta\bar{\mu}_{\alpha}}+\frac{\varepsilon}{2}\frac{\partial U(\varphi)}{\partial u_{\alpha}(\bold{r})}G(\bold{r},\bold{r}|\varphi)\nonumber\\=\Lambda_{\alpha}^{-3}e^{\beta\bar{\mu}_{\alpha}}\left(1-\frac{q_{\alpha}^2}{2(k_{B}T)^2}G(\bold{r},\bold{r}|\varphi)\right)
\end{align}
that yields 
\begin{equation}
\bar{\mu}_{\alpha}=k_{B}T\ln\left(\Lambda_{\alpha}^3\bar{c}_{\alpha}\right)+\frac{q_{\alpha}^2}{2k_{B}T}G(\bold{r},\bold{r}|\varphi).
\end{equation}
Thus, for the bulk, where $u_{\alpha}=0$ and $\varphi=0$, we have
\begin{equation}
\label{mu}
\mu_{\alpha}=k_{B}T\ln\left(\Lambda_{\alpha}^3{c}_{\alpha}\right)+\frac{q_{\alpha}^2}{2k_{B}T}\left(G(0)-G_0(0)\right),
\end{equation}
We subtracted the infinite constant $q_{\alpha}^2G_0(0)/{2k_{B}T}$ from the final result to avoid infinite terms. This can be justified by recognizing that the chemical potentials are determined up to an arbitrary constant. We also considered that in the bulk the Green's function is translation invariant, i.e. $G(\bold{r},\bold{r}^{\prime}|0)=G(\bold{r}-\bold{r}^{\prime})$. The second term in eq. (\ref{mu}) is the fluctuation correction to the ideal gas bulk chemical potential which can be rewritten as follows
\begin{equation}
\label{mu_1}
\mu_{\alpha}^{(1)}=\frac{q_{\alpha}^2}{2k_{B}T}\int\frac{d^2\bold{q}}{(2\pi)^2}\left(G(q|z,z)-G_{0}(q|z,z)\right),
\end{equation}
where $G_{0}(q|z,z^{\prime})= {\exp\left(-q|z-z^{\prime}|\right)}/{2q \beta\varepsilon}$. Further, calculating the integral (\ref{mu_1}) and taking into account that for ideal gas reference system $\kappa=\left(\sum_{\alpha}q_{\alpha}^2c_{\alpha}/\varepsilon k_{B}T\right)^{1/2}$, after some algebra, we arrive at the classical Debye-H{\"u}ckel expression
\begin{equation}
\sigma_{zz}=-P_{o}=-k_{B}T\sum\limits_{\alpha}c_{\alpha}+\frac{k_{B}T\kappa^3}{24\pi},
\end{equation}
where $P_{o}$ is the osmotic pressure of the ions in the bulk.
Performing the same calculations for other nonzero components of the correlation stress tensor, we obtain
\begin{equation}
\sigma^{(cor)}_{xx}=\sigma^{(cor)}_{yy}=\frac{k_{B}T}{4}\int\frac{d^2\bold{q}}{(2\pi)^2}\left(q-\frac{q^2}{\sqrt{q^2+\kappa^2}}-\frac{\kappa^2}{2q}\right)=-\frac{k_{B}T\kappa^3}{12\pi}.
\end{equation}
Taking into account that $\sigma_{xx}^{(0)}=\sigma_{yy}^{(0)}=\sigma_{zz}^{(0)}$, we obtain, as should be, the isotropic stress tensor in the bulk fluid, that is,
$\sigma_{ij}=-P_{o}\delta_{ij}$.

\bibliographystyle{aipnum4-2}
\bibliography{name}

\end{document}